\documentstyle[12pt]{article}
\newcommand{\lsim}{\stackrel{\mbox{\raisebox{-0.1ex}{\scriptsize $<$}}}
{\mbox{\raisebox{-0.5ex}{\scriptsize $\sim$}}}}
\newcommand{\gsim}{\stackrel{\mbox{\raisebox{-0.1ex}{\scriptsize $>$}}}
{\mbox{\raisebox{-0.5ex}{\scriptsize $\sim$}}}}
\newlength{\myleftmargin}
\newlength{\paperwidth}
\begin{document}
\renewcommand{\thefootnote}{\fnsymbol{footnote}}
\begin{flushright}
KOBE--FHD--95--05~\\
June~~~25~~~1995
\end{flushright}

\begin{center}
{\bf Polarized Gluons in the Nucleon}\footnote[1]{Talk
presented by T. Morii at the
Workshop on GeV Scale Physics as a Probe into New Physics,
Toyama, 26-28 June, 1995.} \\
\end{center}
\begin{center}
\vspace{1em}
\underline{T. Morii$^{1, 2}$}, S. Tanaka$^1$ and T. Yamanishi$^2$\\
\end{center}
\vspace{1em}
${}^1$Faculty of Human Development, Kobe University, Nada, Kobe 657, Japan\\
${}^2$Graduate School of Science and Technology,
Kobe University, Nada, Kobe 657, Japan

\vspace{1em}
\begin{center}
{\bf Abstract}
\end{center}

QCD suggests that gluons in the nucleon play an important role
in {\it so-called} ``the proton spin problem''.
In this talk, the behavior
of the polarized gluon distribution in the nucleon is discussed by
using the positivity condition of distribution functions together
with the unpolarized and polarized experimental data.

\parskip 1em

\noindent
{\bf 1. Introduction}

The proton spin problem issured from the measurement
of the polarized structure function of proton $g_1^p(x)$ by the
EMC group\cite{EMC}, has caused a great interest in particle and nuclear
physics community. The EMC\cite{EMC} and SMC\cite{SMC94}
data on $g_1^p(x)$ suggest
that very little of the proton spin is carried by quarks
($\Delta \Sigma \approx 0$) and furthermore the polarized $s$-quark
density is rather large ($\Delta s \approx -0.12$\cite{SMC94}).  The result is
apparently incompatible with the prediction of the naive quark
and/or parton model.  So far a number of theoretical approaches
have been proposed to solve the problem. Among these approaches,
there is an interesting idea that gluons contribute significantly
to the proton spin through the $U_A(1)$ anomaly of QCD\cite{anomaly}.
In this model the polarized quark densities taking part in
$g_1^p$ should be modified as $\Delta q\rightarrow
\Delta q-\frac{\alpha _s}{2\pi}\Delta G$, where $\Delta q$ and
$\Delta G$ are polarized quark and gluon densities, respectively,
and one can reproduce the EMC/SMC data quite well by taking
rather large $\Delta G(\approx 5\sim 6)$\cite{Kobayakawa92}.  In addition,
such a large $\Delta G$ can solve an apparent discrepancy
on the magnitude of $\Delta s$ between rather large $\Delta s$
derived from EMC/SMC data and a restrictive bound
$|\Delta s| \leq 0.05^{+0.02}_{-0.05}$\cite{Preparata88}
derived from the experiment on charm productions in neutrino
deep inelastic scatterings\cite{CDHS}.
These observations tell us that it is
very important to determine the behavior of the polarized gluon
distribution $\delta G(x)$ and its magnitude
$\Delta G(=\int^1_0 \delta G(x)dx)$.

\noindent
{\bf 2. Polarized gluon distributions}

In this talk, we discuss the $x$ dependence of the polarized
gluon distribution $\delta G(x)$.
Let us start with the functional form of $\delta G(x)$.
By taking account of the plausible behavior of
$\delta G(x)$ near $x\approx 0$ and $x\approx 1$, we assume
\begin{eqnarray}
\delta G(x) = G_+(x) - G_-(x)=B~x^{\gamma}~(1-x)^p~(1+C~x)~,
\label{eqn:delta G}
\end{eqnarray}
where $G_+(x)$ and $G_-(x)$ are the gluon distributions with helicity parallel
and antiparallel to the proton helicity, respectively. We further assume for
simplicity $G_+(x)\approx G_-(x)$ at large $x$ and take $C=0$. Then we
have two unknown parameters, $\gamma$ and $p$. $B$ is determined from the
normalization, $\Delta G=5.32$\cite{Morii},
which is derived from the SMC data.
We are interested in the behavior of
$\delta G(x)$
under the condition of large $\Delta G$ ($=5.32$) and search the allowable
region of $\gamma$ and $p$. In order to do this,
we require the positivity condition of distribution functions and utilize
the recent results of several polarization experiments.

\noindent
{\bf 3. Restriction on the $x$ dependence of $\delta G(x)$}

As a preliminary, to examine the behavior of $\delta G(x)$ in
eq.(\ref{eqn:delta G}) for various values of $\gamma$ and $p$,
$\gamma$ is varied from $-0.9$ to $0.3$ at intervals of $0.3$
while $p$ is chosen
independently as $5$, $10$, $15$, $17$ and $20$.

(i) First, let us consider the positivity condition
of distribution functions to
restrict $\gamma$ and $p$. In the same way as in eq.(\ref{eqn:delta G})
for the polarized gluon distribution,
the unpolarized gluon distribution $G(x)$ is assumed as
\begin{eqnarray}
G(x) = G_+(x) + G_-(x)=\frac{A}{x^{\alpha}}~(1-x)^k.
\label{eqn:G(x)}
\end{eqnarray}
 Since $G_+(x)$ and $G_-(x)$
are both positive, we obtain from eqs.(\ref{eqn:delta G})
and (\ref{eqn:G(x)})\cite{Morii}
\begin{equation}
|~\Delta G~| \leq \frac{\Gamma (\gamma +1)~\Gamma (p+1)~\Gamma (k+3-\alpha)~
                 (\alpha +\gamma+p-k)^{\alpha +\gamma+p-k}}
          {\Gamma (\gamma +p+2)~\Gamma (k+1)~\Gamma (2-\alpha)~
          (\alpha +\gamma)^{\alpha +\gamma}(p-k)^{p-k}}\int^1_0 xG(x)dx~.
\label{eqn:bound}
\end{equation}
To restrict the region of $\gamma$ and $p$ from
this inequality (\ref{eqn:bound}) with
$\Delta G=5.32$, we need to know the value of $\alpha$ and $k$ in $G(x)$ and
the intergral value of $xG(x)$ as well. As for the $x$ dependence of $G(x)$,
using  experimental data of J/$\psi$ productions for unpolarized muon--nucleon
scatterings\cite{NMC91,EMC92}, we have two possible types of parameterization
of $G(x)$ at $Q^2\simeq M_{J/\psi}^2$ GeV$^2$(see Fig.1),
\begin{eqnarray}
Type~A~~~~~~~~~G(x) &=& 3.35~\frac{1}{x}~(1-x)^{5.7}~,
\label{eqn:type-a}\\
Type~B~~~~~~~~~G(x) &=& 2.36~\frac{1}{x^{1.08}}~(1-x)^{4.65}~.
\label{eqn:type-b}
\end{eqnarray}
For Type A, $\alpha$ is taken to be $1$ by considering the ordinary Pomeron P,
and parametrized so as to fit the data. On the other hand, $\alpha$ is chosen
to be $1.08$ in Type B which is recently derived from the analysis of the
experimental data of the total cross section\cite{Donnachie}.
 Inserting these parameters into inequality
(\ref{eqn:bound}) with $\Delta G=5.32$, the allowed regions of $\gamma$ and
$p$ are obtained(see Table 1 and Fig.2).
In Fig.2, the region below solid
or dashed lines is excluded by (\ref{eqn:bound}).
We see that a wide region of $\gamma$ and $p$ which
satisfies the SMC data and the positivity condition simultaneously, is
allowable.

(ii) Secondly, to restrict further the allowable region of $\gamma$ and $p$, we
compare our model calculations with the two--spin asymmetries
$A^{\pi^0}_{LL}(\stackrel{\scriptscriptstyle(-)}{p}\stackrel{}{p})$ for
inclusive $\pi^0$--productions measured by E581/704 Collaboration using
polarized (anti-)proton beams and polarized proton targets\cite{E581}.
Taking $\delta G(x)$ with the sets of $(\gamma, p)$
allowed by the criterion of positivity, we calculate numerically
$A^{\pi^0}_{LL}(\stackrel{\scriptscriptstyle(-)}{p}\stackrel{}{p})$(see
Table 2 and Fig.3).
In doing so, polarized quark distributions
$\delta q(x) (q=u,d,s)$ are necessary and taken from
ref.\cite{Kobayakawa92}.  In order to reproduce the experimental data,
$x\delta G(x)$ must have a peak at a smaller $x$ than
$0.05$ and decrease very rapidly with $x$.

(iii)~~Finally, we look into the spin--dependent structure function of proton
$g_1^p(x)$\cite{SMC94} and that of deuteron $g_1^d(x)$\cite{SMC93}.
The results calculated by using
our $\delta q(x)$ and $\delta G(x)$ with $(\gamma, p)$ having survived
the criteria of cases (i) and (ii) are shown in Fig.4. Furthermore,
to examine the adequacy of calculations more objectively,
$\chi^2$/DOF and the confidence level for $xg_1^p(x)$ and $xg_1^d(x)$
are calculated and shown in Table 3.

\noindent
{\bf 4. Conclusion and discussion}

Within the models with large $\Delta G$ ($=5.32$), we have studied
the shape of the polarized gluon distribution. By using the positivity
condition of distribution functions together with the experimental data on
the two--spin asymmetries
$A^{\pi^0}_{LL}(\stackrel{\scriptscriptstyle(-)}{p}\stackrel{}{p})$
and the spin--dependent structure functions of $g_1^p(x)$ and $g_1^d(x)$,
we have restricted the $x$ dependence of $\delta G(x)$ given in
eq.(\ref{eqn:delta G}). As for the values of $\gamma$ and $p$,
$\gamma\lsim -0.3$ and $p\gsim 10$ seem favorable in our analysis.
If $\gamma$ and $p$ are fixed in this region, for example,
as $\gamma=-0.6$ and $p=17$,
\begin{eqnarray}
\delta G(x) = 9.52~x^{-0.6}~(1-x)^{17}~,
\label{eqn:delta G(x)}
\end{eqnarray}
one can reproduce
all existing data quite successfully. Needless to say, with
large $\Delta G$ ($=5.32$) the $\Delta s$
becomes small and can be reconciled with the bound
$|\Delta s| \leq 0.05^{+0.02}_{-0.05}$.
In the Regge terminology,
$\gamma=-0.6$ happens to be closer to the one for unpolarized
valence quark distributions rather than for unpolarized gluon
distributions\cite{Regge}, and $p$ seems to be inconsistent with the
prediction of counting rules\cite{count}. At present, we do not
know the theoretical ground on the origin of
these values of $\gamma$ and $p$.
 Furthermore, if $\Delta G$ is so large
($\approx 5\sim 6$), we are to have an approximate relation
$\langle L_Z\rangle_{q+G}\approx -\Delta G$ from the proton spin sum rule,
$\frac{1}{2}=\frac{1}{2}\Delta\Sigma+\Delta G+\langle L_Z\rangle_{q+G}$,
where $\frac{1}{2}\Delta\Sigma$ represents the sum of the spin carried by
quarks. Unfortunately, nobody knows the underlying physics of it. These are
still problems to be solved even though the idea of the U$_A$(1) anomaly is
attractive.

It is interesting to comment on another approach which has mentioned to this
problem.
Recently, Brodsky, Burkardt and Schmidt (BBS)\cite{Brod94} have proposed an
interesting model of the polarized gluon distribution which incorporates
color coherence and the counting rule at small and large $x$.
At $x\approx 0$, the color coherence argument gives $\delta G(x)/G(x)\approx
\frac{x}{3}\langle\frac{1}{y}\rangle$ with $\langle\frac{1}{y}\rangle\simeq 3$,
where $\langle\frac{1}{y}\rangle$ presents the first inverse moment of the
quark light--cone momentum fraction distributions in the lowest Fock state of
the proton, and leads to a relation $\gamma=-\alpha +1$\cite{Brod94,Gehrmann}.
Then, contrary to our result, $\gamma\lsim -0.3$, they have taken
$\gamma=0$ by choosing $\alpha=1$ which is an ordinary Pomeron intercept value.
 Although the integrated
value of $\delta G(x)$ in the BBS model is small such as $\Delta G=0.45$,
the model reproduces well the EMC data $g_1^p(x)$, $g_1^n(x)$ and
$g_1^d(x)$. In addition, we calculated
$A^{\pi^0}_{LL}(\stackrel{\scriptscriptstyle(-)}{p}\stackrel{}{p})$
by using the BBS model and found that the model could reproduce
$A^{\pi^0}_{LL}(\bar pp)$ while the predicted value of $A^{\pi^0}_{LL}(pp)$
slightly deviated from the data\cite{Morii94}. The BBS model which has
small $\Delta G$ ($=0.45$) seems to be an alternative to our model with
large $\Delta G$ ($=5.32$), though in the BBS model
$\Delta s$ becomes large and cannot be reconciled with the bound
$|\Delta s| \leq 0.05^{+0.02}_{-0.05}$.

It is very important to know the behavior of
$\delta G(x)$ and $\delta s(x)$ independently
in order to understand the spin structure of the nucleon.
However the polarization
experiments are still in their beginning and the form of these functions
is not yet clear. We hope they will be determined in the
forthcoming experiments.

\vfill\eject

\begin{center}
\begin{tabular}[t]{|c|c|c|c|c|c|c|}\hline
\multicolumn{2}{|c|}{ }& \multicolumn{5}{c|}{p}\\
\cline{3-7}
\multicolumn{2}{|c|}{ } & 5 & 10 & 15 & 17 & 20 \\ \hline
\multicolumn{1}{|c|}{ } & 0.3& $\ast$ & $\times$ & $\times$ & $\times$ &
$\times$ \\
\cline{2-7}
\multicolumn{1}{|c|}{ } & 0.0& $\ast$ & $\times$ & $\bigcirc$ & $\bigcirc$ &
$\bigcirc$ \\
\cline{2-7}
$\gamma$ & $-0.3$ & $\ast$ & $\bigcirc$ & $\bigcirc$ & $\bigcirc$ & $\bigcirc$
\\
\cline{2-7}
\multicolumn{1}{|c|}{ } & $-0.6$ & $\ast$ & $\bigcirc$ & $\bigcirc$ &
$\bigcirc$ & $\bigcirc$ \\
\cline{2-7}
\multicolumn{1}{|c|}{ } & $-0.9$ & $\ast$ & $\bigcirc$ & $\bigcirc$ &
$\bigcirc$ & $\bigcirc$ \\ \hline
\end{tabular}
\hspace{2cm}
\begin{tabular}[t]{|c|c|c|c|c|c|c|}\hline
\multicolumn{2}{|c|}{ }& \multicolumn{5}{c|}{p}\\
\cline{3-7}
\multicolumn{2}{|c|}{ } & 5 & 10 & 15 & 17 & 20 \\ \hline
\multicolumn{1}{|c|}{ } & 0.3& $\times$ & $\times$ & $\times$ & $\times$ &
$\times$ \\
\cline{2-7}
\multicolumn{1}{|c|}{ } & 0.0& $\times$ & $\times$ & $\times$ & $\bigcirc$ &
$\bigcirc$ \\
\cline{2-7}
$\gamma$ & $-0.3$ & $\times$ & $\bigcirc$ & $\bigcirc$ & $\bigcirc$ &
$\bigcirc$ \\
\cline{2-7}
\multicolumn{1}{|c|}{ } & $-0.6$ & $\times$ & $\bigcirc$ & $\bigcirc$ &
$\bigcirc$ & $\bigcirc$ \\
\cline{2-7}
\multicolumn{1}{|c|}{ } & $-0.9$ & $\bigcirc$ & $\bigcirc$ & $\bigcirc$ &
$\bigcirc$ & $\bigcirc$ \\ \hline
\end{tabular}
\end{center}

\begin{description}
\item[Table 1:] The various sets of $\gamma$ and $p$ which we have
examined. The circles denote the sets allowed from (\ref{eqn:bound})
whereas the crosses present the ones excluded from (\ref{eqn:bound}).
The left--side (right--side) table corresponds to Type A (Type B) of $G(x)$.
In the left--side table, the sets of $(\gamma, p)$ with $p=5$, which are
filled with asterisks, are not used for calculations because $p-k$ should be
larger than zero.
\end{description}

\parskip 2em

\begin{center}
\begin{tabular}[t]{|c|c|c|c|c|c|c|}\hline
\multicolumn{2}{|c|}{ }& \multicolumn{5}{c|}{p}\\
\cline{3-7}
\multicolumn{2}{|c|}{ } & 5 & 10 & 15 & 17 & 20 \\ \hline
\multicolumn{1}{|c|}{ } & 0.3& $-$ & $-$ & $-$ & $-$ & $-$ \\
\cline{2-7}
\multicolumn{1}{|c|}{ } & 0.0& $-$ & $-$ & $\times$ & $\times$ & $\bigcirc$ \\
\cline{2-7}
$\gamma$ & $-0.3$ & $-$ & $\times$ & $\times$ & $\bigcirc$ & $\bigcirc$ \\
\cline{2-7}
\multicolumn{1}{|c|}{ } & $-0.6$ & $-$ & $\times$ & $\bigcirc$ & $\bigcirc$ &
$\bigcirc$ \\
\cline{2-7}
\multicolumn{1}{|c|}{ } & $-0.9$ & $\times$ & $\bigcirc$ & $\bigcirc$ &
$\bigcirc$ & $\bigcirc$ \\ \hline
\end{tabular}
\hspace{2cm}
\begin{tabular}[t]{|c|c|c|c|c|c|c|}\hline
\multicolumn{2}{|c|}{ }& \multicolumn{5}{c|}{p}\\
\cline{3-7}
\multicolumn{2}{|c|}{ } & 5 & 10 & 15 & 17 & 20 \\ \hline
\multicolumn{1}{|c|}{ } & 0.3& $-$ & $-$ & $-$ & $-$ & $-$ \\
\cline{2-7}
\multicolumn{1}{|c|}{ } & 0.0& $-$ & $-$ & $\bigcirc$ & $\bigcirc$ & $\bigcirc$
\\
\cline{2-7}
$\gamma$ & $-0.3$ & $-$ & $\times$ & $\bigcirc$ & $\bigcirc$ & $\bigcirc$ \\
\cline{2-7}
\multicolumn{1}{|c|}{ } & $-0.6$ & $-$ & $\bigcirc$ & $\bigcirc$ & $\bigcirc$ &
$\bigcirc$ \\
\cline{2-7}
\multicolumn{1}{|c|}{ } & $-0.9$ & $\times$ & $\bigcirc$ & $\bigcirc$ &
$\bigcirc$ & $\bigcirc$ \\ \hline
\end{tabular}
\end{center}

\begin{description}
\item[Table 2:] The various sets of $\gamma$ and $p$ which we have
examined. The circles denote the sets allowed from the $A_{LL}^{\pi^0}$
whereas the crosses present the ones excluded from it.
The minuses denote the sets excluded from Table 1.
The left--side (right--side) table corresponds to the $A_{LL}^{\pi^0}$ for
$pp$ collisions ($\bar pp$ collisions).
\end{description}

\vfill\eject

\begin{center}
{\Large (A)}\\
\begin{tabular}[t]{|c|c|c|c|c|c|}\hline
\multicolumn{2}{|c|}{ }& \multicolumn{4}{c|}{p}\\
\cline{3-6}
\multicolumn{2}{|c|}{ } & 10 & 15 & 17 & 20 \\ \hline
\multicolumn{1}{|c|}{ } & 0.0& $-$ & $-$ & $-$ &
 \begin{tabular}{c}
   $0.766$ \\
   $(83.5\%)$
 \end{tabular} \\
\cline{2-6}
$\gamma$ & $-0.3$ & $-$ & $-$ &
 \begin{tabular}{c}
   $0.501$ \\
   $(99.3\%)$
 \end{tabular} &
 \begin{tabular}{c}
   $0.444$ \\
   $(99.8\%)$
 \end{tabular} \\
\cline{2-6}
\multicolumn{1}{|c|}{ } & $-0.6$ & $-$ &
 \begin{tabular}{c}
   $0.444$ \\
   $(99.8\%)$
 \end{tabular} &
 \begin{tabular}{c}
   $0.448$ \\
   $(99.8\%)$
 \end{tabular} &
 \begin{tabular}{c}
   $0.474$ \\
   $(99.6\%)$
 \end{tabular} \\
\cline{2-6}
\multicolumn{1}{|c|}{ } & $-0.9$ &
 \begin{tabular}{c}
  $0.646$ \\
  $(94.5\%)$
 \end{tabular} &
 \begin{tabular}{c}
  $0.728$ \\
  $(87.7\%)$
 \end{tabular} &
 \begin{tabular}{c}
  $0.771$ \\
  $(82.9\%)$
 \end{tabular} &
 \begin{tabular}{c}
  $0.839$ \\
  $(73.3\%)$
 \end{tabular} \\ \hline
\end{tabular}
\end{center}

\begin{center}
{\Large (B)}\\
\begin{tabular}[t]{|c|c|c|c|c|c|}\hline
\multicolumn{2}{|c|}{ }& \multicolumn{4}{c|}{p}\\
\cline{3-6}
\multicolumn{2}{|c|}{ } & 10 & 15 & 17 & 20 \\ \hline
\multicolumn{1}{|c|}{ } & 0.0& $-$ & $-$ & $-$ &
 \begin{tabular}{c}
   $1.006$ \\
   $(43.8\%)$
 \end{tabular} \\
\cline{2-6}
$\gamma$ & $-0.3$ & $-$ & $-$ &
 \begin{tabular}{c}
   $0.726$ \\
   $(71.5\%)$
 \end{tabular} &
 \begin{tabular}{c}
   $0.629$ \\
   $(80.6\%)$
 \end{tabular} \\
\cline{2-6}
\multicolumn{1}{|c|}{ } & $-0.6$ & $-$ &
 \begin{tabular}{c}
   $0.539$ \\
   $(87.9\%)$
 \end{tabular} &
 \begin{tabular}{c}
   $0.501$ \\
   $(90.4\%)$
 \end{tabular} &
 \begin{tabular}{c}
   $0.463$ \\
   $(92.7\%)$
 \end{tabular} \\
\cline{2-6}
\multicolumn{1}{|c|}{ } & $-0.9$ &
 \begin{tabular}{c}
   $0.581$ \\
   $(84.7\%)$
 \end{tabular} &
 \begin{tabular}{c}
   $0.548$ \\
   $(87.2\%)$
 \end{tabular} &
 \begin{tabular}{c}
   $0.547$ \\
   $(87.2\%)$
 \end{tabular} &
 \begin{tabular}{c}
   $0.554$ \\
   $(86.7\%)$
 \end{tabular} \\ \hline
\end{tabular}
\end{center}

\begin{description}
\item[Table 3:] The values of $\chi^2$/DOF and the confidence level
(in the parentheses) for $(\gamma, p)$
which have survived in Table 2. (A) ((B))
corresponds to the $xg_1^p(x)$ ($xg_1^d(x)$). The DOF of $\chi^2$ to
$xg_1^p(x)$ ($xg_1^d(x)$) is $34$ ($11$). The minus signs denote the
sets excluded from Tables 1 and 2.
\end{description}

\vfill\eject

\begin{center}
{\large \bf Figure captions}
\end{center}
\begin{description}

\item[Fig. 1:] The parametrization of the gluon distribution
functions $xG(x, Q^2)$ at $Q^2\approx 10$GeV$^2$. The solid (dashed) line
denotes Type A (B). The data of opened (closed) circles are taken from
\cite{NMC91} (\cite{EMC92}).

\parskip 1em

\item[Fig. 2:] The allowed region by (\ref{eqn:bound}) for $\gamma$ and $p$.
The solid (dashed) line corresponds to Type A (B). The region below the
lines are excluded.

\parskip 1em

\item[Fig. 3:] The produced $\pi^0$ transverse momenta $p_T$ dependence of
$A^{\pi^0}_{LL}(\stackrel{\scriptscriptstyle(-)}{p}\stackrel{}{p})$ for
various $p$ ($=5 - 20$) with (a) $\gamma=0$ and (b)
$\gamma=-0.9$. Data are taken from \cite{E581}.

\parskip 1em

\item[Fig. 4:] The $x$ dependence of $xg_1^p(x)$ and $xg_1^d(x)$
for various $p$ ($=5 - 20$) with (a) $\gamma=0$ and (b)
$\gamma=-0.9$. The data of $xg_1^p(x)$ ($xg_1^d(x)$) are taken from
\cite{SMC94,EMC} (\cite{SMC93}).
\end{description}
\end{document}